\documentstyle[preprint,aps]{revtex}
\widetext
\draft
\tighten

\begin{document}

\title{Convection displacement current and alternative form of
Maxwell-Lorentz equations}

\bigskip

\author{{\bf Andrew E. Chubykalo and Roman Smirnov-Rueda}
\thanks{Instituto de Ciencia de Materiales, C.S.I.C., Madrid, Spain}}

\address {Escuela de F\'{\i}sica, Universidad Aut\'onoma de Zacatecas \\
Apartado Postal C-580\, Zacatecas 98068, ZAC., M\'exico}

\date{\today}

\maketitle


\baselineskip 7mm
\bigskip
\bigskip
\bigskip
\bigskip


\clearpage
\begin{abstract}
Some mathematical inconsistencies in the conventional form of Maxwell's
equations extended by Lorentz for a single charge system are discussed.
To surmount these in framework of Maxwellian theory,
a novel convection displacement current is considered as
additional and complementary to the famous Maxwell displacement
current. It is shown that this form of the
Maxwell-Lorentz equations is similar to that proposed by Hertz for
electrodynamics of bodies in motion. Original Maxwell's equations
can be considered as a valid approximation for a continuous and closed (or
going to infinity) conduction current. It is
also proved that our novel form of the Maxwell-Lorentz equations is
relativistically invariant. In particular, a relativistically invariant
gauge for quasistatic fields has been found to replace the
non-invariant Coulomb gauge. The new gauge condition contains the famous
relationship between electric and magnetic potentials for one uniformly
moving charge that is usually attributed to the Lorentz
transformations. Thus, for the first time, using the convection
displacement current, a physical interpretation is given to the
relationship between the components of the four-vector of quasistatic
potentials. A rigorous application of the new gauge
transformation with the Lorentz gauge transforms the
basic field equations
into an independent pair of differential equations responsible for
longitudinal and transverse fields, respectively. The
longitudinal components can be interpreted
exclusively from the standpoint of the instantaneous
"action at a distance" concept and leads to necessary conceptual
revision of the conventional Faraday-Maxwell field.
The concept of electrodynamic dualism is
proposed for self-consistent classical electrodynamics. It
implies simultaneous coexistence of instantaneous long-range
(longitudinal) and Faraday-Maxwell short-range (transverse) interactions
that resembles in this aspect the basic idea of Helmholtz's
electrodynamics.

\end{abstract}

\section{Introduction}
In the early part of 19'th century, the highest importance was attached to
electricity and magnetism in an attempt to justify a particular outlook
on the world. After Faraday's fundamental discovery of the law of
induction, the challenge presented itself of unifying
electrodynamics into a coherent
whole out of the electrostatics of Neumann and
magnetostatics of Amp\`ere. From
the historical point of view it was an exiting and
unique moment in the
development of physics [1]. Two rival concepts: instantaneous
action at a
distance and Faraday's field concept were waiting for decisive
progress in theory and experiment: to be confirmed or rejected. The
state of
electromagnetism was characterized
by the search for a correct and unambiguous concept by excluding all
alternatives.
      In 1848, Wilhelm Weber was the first to attempt to
unify electromagnetic theory [2]. At the time, Weber's action
at a distance type theory was felt to be a great advance because it
reproduced the results of both Neumann and Amp\`ere using an
 analytical treatment of induced currents. Some years
later in 1855-56, James Clerk Maxwell communicated to the Cambridge
Philosophical Society his first memoir: an attempt to develop a
comprehensive mechanical conception of the electromagnetic field. This
afterwards developed into a powerful field theory [3,4]. In the years
immediately following the Maxwell's {\it Treatise}, a certain amount of
evidence in favor of his theory was furnished by experiment.
Nevertheless,
it was not sufficient to make a choice
between the two alternatives. Two
theories; those of Weber and Maxwell, could satisfactory describe
the major known
electromagnetic phenomena in spite of many internal limitations.

      In 1870, in consequence of long and fruitless opposition of two
rival concepts in electromagnetism, and for the purpose
of reconciling
them, a compromise theory was proposed by Helmholtz [5].
Soon it became the accepted theory in Germany and
continental Europe. In particular, Hertz and Lorentz
got to know
the Maxwellian theory through Helmholtz's equations.
The compromise theory
of Helmholtz also was incomplete and internally inconsistent.
It applied exclusively to a  dielectric medium at rest
and did not take into consideration displacement currents in vacuo
(ether). Helmholtz's theory allowed simultaneous coexistence of
instantaneous longitudinal electric modes and transverse
electric and magnetic waves propagating with a
certain finite spread
velocity. The magnetic field, however, could be transmitted
only through transverse
modes. There was conceptual conflict with Maxwell's theory.
For instance, the value of transverse wave velocity
differed considerably
from that predicted by the Maxwell field equations
in a dielectric medium.

At that point in time there was an urgent need for reliable
new experimental data. The Berlin Academy proposed as a prize subject
 ``To establish
 experimentally a relation between electromagnetic action and the
 polarization of dielectrics". This investigation led Hertz to a
  discovery of great importance that is now widely accepted as a crucial
 moment in the history of electromagnetism. Hertz succeeded in directly
 observing the propagation of electromagnetic waves in free space with
 the velocity predicted by Maxwell and thereby decisively tested his
 theory. Another two alternatives (Weber's and Helmholtz's theories) had
 been rejected in spite of the fact that Helmholtz's theory, corrected
 for the spread velocity for transverse modes no longer
 contradicted the Hertz experiment. In the literature devoted to the
 history and
methodology of physics there was great discussion whether one
 experiment such as of Hertz could be considered as a decisive argument
 at the choice between a few alternatives [6]. From here on, all main
 investigations in electromagnetism were based on
 Maxwell's equations. Nevertheless, this theory still suffered from
 some short comings inherent to
 its predecessors. To be more specific, Maxwell's equations for steady
 state processes could be compatible only with continuous,
 closed currents derived from
 Amp\`ere's magnetostatics. The theory
 needed to be extended to a one charge system.

In 1881, the first examination of the matter from the standpoint of
 Maxwell's theory was undertaken by J.J. Thompson [7] and later by O.
 Heaviside [8]. They found that the change in the location of the charge
 (they considered an electrostatically charged body in motion)
 must produce a
 continuous alteration of the electric field at any point in the
 surrounding medium (the point of difference with Maxwell's theory of
 stationary fields). In the language adopted by Maxwell, there must be
 displacement currents in the vacuum attributed to the magnetic effects of
 moving charge. Developing this approach, FitzGerald pointed out in
 a short
 but valuable note [9] that Thompson's method must be identified with the
 basic hypothesis of Maxwell's theory about the total current. This
 conclusion was based on the assumption that a moving charge itself was
 to be counted as a current element. Then the total current, thus composed
 of the displacement currents and a moving charge, was circuital, in
 accordance with Maxwell's fundamental ideas.

 As the result of their investigations, Thompson and Heaviside found the
electric and magnetic fields of a moving charge that now can be verified
directly by use of Lorentz transformations. They also gave for the first
time an explicit formula for mechanical force acting on an electric
charge which is moving in a magnetic field, now known as the Lorentz
force. It was the first successful demonstration that Maxwell's theory
might be extended to a one charge system.

In spite of the advances which were effected by Maxwell and his earliest
followers a more general theory was needed for bodies in motion.
In 1890, Hertz made the first attempt to build up such a theory. As in
all 19'th century methods, Hertz based his considerations on
an adequate model of ether. As Hertz himself expressed it, ``the ether
contained within ponderable bodies moved with them". Like Maxwell, he
assumed that the state of the compound system (matter and
ether) could be specified in the same way when matter was in motion or
at rest. Thus, it was Hertz who established for the first time
the relativity principle in electrodynamics. It could not have been as
general as the Einstein-Poincar\`e relativity principle since it was
tied to a model of the ether. However, as will be shown
in this work,  Hertz's equations, carefully reconsidered,
 can be treated as a
relativistically invariant and alternative form of Maxwell's equations.
The results of Hertz's theory resembled in many respects those of
Heaviside who likewise was disposed to accept an additional term in
Maxwell's equations involving $curl({\bf V}\times{\bf E})$ that they
called the {\it current of dielectric convection}.

However, contemporaries of Hertz did not accept as altogether successful
his
attempts to extend the theory of the electromagnetic field to the case in
which ponderable bodies are in motion. Hertz's assumptions on the
ether were conceptually inconsistent with the existing interpretation of
Fizeau's experiment and seemed to disagree with Fresnel's formulae.
Meanwhile, in the decade of 1890, the
electrodynamics of moving media was systematically treated by Lorentz.
The principal differences by which the theory advanced by Lorentz was
distinguished from Hertz's one lay in the conception of "electron" and in
the model of the ether. Lorentz designed his equations in accord with the
successful Fresnel theory. A distinction had been made between matter
and ether by assuming that a moving ponderable body could not communicate
its motion to the ether which surrounds it. This hypothesis corresponded
to the ether in rest and implied that no part of the ether could be in
motion relative to any other part. The correctness of Lorentz's hypothesis,
as opposed to that of Hertz's one, was afterwards confirmed by various
experiments, one of them was the discovery of the electron. The
experimental
confirmation that electric charges resided in atoms put the Lorentz
theory at the center of scientific interest and made it the basis for all
further investigations in the area of electromagnetism. The
reconciliation of the electromagnetic theory with Fresnel's law of the
propagation of light in moving bodies achieved in the framework of
Lorentz's theory was considered as a distinct advance. Nevertheless, the
theory was far from complete and posed various problems. Some of them
were internal, others were related to the direct application
of the theory.

The decisive consolidation of the Lorentz theory of bodies in
motion was made by A. Einstein who established a fundamental
relativity principle. Since then a more comprehensive and consistent
scheme of electromagnetic phenomena has been available.
This was a turning-point in the development of electromagnetism. The
mechanistic conception of ether had been abolished. Fresnel's formulae
could be treated now on the basis of the relativistic law of addition of
velocities. The principal arguments against
the Hertz's theory were removed.
However, Lorentz's theory, conceptually renovated,
(without any attempt to do the same with Hertz's theory) was
retained as the basis of conventional classic electrodynamics. In spite of all
advances, the electromagnetic theory of Lorentz has internal
inconsistencies such as the self-reaction force
({\it self-interaction}), infinite contribution of self-energy, the
concept of electromagnetic mass, indefiniteness in the flux of
electromagnetic energy, unidirectionality of radiation phenomena with
respect to motion reversal in the basic Maxwell equations etc.
The advent of
quantum mechanics in the early part
of this century brought the hope that
all classical difficulties could be straightened out in the
framework of
quantum electrodynamics. Quantum field theory does alleviate some
problems but cannot surmount them without introducing unjustified
renormalization methods. The principal difficulties in
Maxwell's theory persist and do not disappear in spite of taking into
account quantum mechanical modifications. The Hertz theory, renovated
in accordance with Einstein's
relativity principle, might provide a more satisfactory alternative.

As will be shown here, a modification of the Maxwell-Hertz equations can
be made equivalent to a form of the Maxwell-Lorentz
 set of equations,
and their rigorous solution shows the existence of a longitudinal,
explicitly time independent, component of the electromagnetic field.
This
finding supports some recent investigations in the new area of
longitudinal
field solutions of the fundamental electromagnetic field equations
proposed by M. W. Evans and J.-P. Vigier [10,11].

The general solution of our modified field equations reproduces
the ``separation of potentials" theory proposed recently in [12] as a
method of removing
the above-mentioned difficulties from classical electromagnetic
theory. In [12] the conventional Faraday-Maxwell
concept of the field was shown to be not fully adequate for Maxwell's
equations. Electrodynamical dualism was developed, and
implies the coexistence of the
instantaneous long-range and Faraday-Maxwell short-range
interactions. For the first time since the Hertz's experimental
discovery, an argument has been developed
in favor of Helmholtz's alternative theory
renovated in the framework of Maxwell's equations.
The concept of action at a distance in this theory differs completely
the action at a distance theories of
postspecial relativity [13,14], theories which assume that only
delayed action at a distance with the speed of light can be consistent
with relativity. The new approach developed in [12]
demonstrates the compatibility of instantaneous action at a
distance with relativistic classical electrodynamics.
The new ``dualism concept" can bridge the gap
between classical and quantum physics. From this point of view, the
instantaneous action at a distance might become a classical analogy of
non-locality in quantum theories.

Let us conclude here this historical background and go on to the analysis
of some difficulties in the Lorentz electrodynamics.

\section{Maxwell-Lorentz equations. Paradox}

Let us write Maxwell's equations for the reference system at rest, grouping
them into two pairs [15]:

\begin{eqnarray}
&& div\,{\bf E}=4\pi\varrho,\\
&& div\,{\bf B}=0
\end{eqnarray}
and
\begin{eqnarray}
&& curl\,{\bf H}=\frac{4\pi}{c}{\bf j}+\frac{1}{c}\frac{\partial{\bf
E}}{\partial t},\\
&& curl\,{\bf E}=-\frac{1}{c}\frac{\partial{\bf
B}}{\partial t}
\end{eqnarray}
at the same time with the continuity equation:
\begin{equation}
\frac{\partial\varrho}{\partial t}+div\,{\bf j}=0.
\end{equation}

In the phenomenological theory of electromagnetism the hypothesis about
the continuous nature of medium was one of the foundations of Maxwell's
speculations. This point of view succeeded in uniting all electromagnetic
phenomena without the necessity to consider a specific structure of
matter. Nevertheless, the macroscopic character of the charge conception
defines all well-known limitations on Maxwell's theory. For instance, the
system of electromagnetic fields equations (1-4) in a steady state
corresponds to a quite particular case of continuous and closed conduction
currents (motionless as a whole).

In 1895, the theory was extended by Lorentz for a system of charges moving
in vacuo. Since then it has been widely assumed that the same basic laws
are valid microscopically and macroscopically in the form of original
Maxwell's equations. It means that in Lorentz form all macroscopic values
of charge and current densities must be substituted by its microscopic
values. Let us write explicitly the Lorentz field equations for one point
charged particle moving in vacuo [15]:

\begin{eqnarray}
&& div\,{\bf E}=4\pi q \delta({\bf r}-{\bf r}_q(t)),\\
&& div\,{\bf B}=0
\end{eqnarray}
and
\begin{eqnarray}
&& curl\,{\bf H}=\frac{4\pi}{c}q{\bf V}\delta({\bf r}-{\bf
r}_q(t))+\frac{1}{c}\frac{\partial{\bf E}}{\partial t},\\
&& curl\,{\bf E}=-\frac{1}{c}\frac{\partial{\bf
B}}{\partial t}
\end{eqnarray}
here ${\bf r}_q(t)$ is the coordinate of the charge at an instant $t$.

In order to achieve a complete description of a system consisting of fields
and charges in the framework of electromagnetic theory, Lorentz
supplemented (6)-(9) by the equation of motion of a particle:
\begin{equation}
\frac{d{\bf p}}{dt}=q{\bf E}+\frac{q}{c}({\bf V}\times{\bf B})
\end{equation}
where ${\bf p}$ is the momentum of the particle.

The equation of motion (10) introduces an expression for the mechanical
force known as Lorentz force which in electron theory formulated by
Lorentz clearly has an axiomatic and empirical character. Later on
we shall discuss some disadvantages related with the adopted status of the
Lorentz force conception.

The macroscopic Maxwell's equations (1)-(4) may be obtained now from
Lorentz's equations (6-9) by some statistical averaging process, using
the structure of material media. The mathematical language for eqs. (6)-(9)
is widely adopted in the conventional classical electrodynamics.

However, there is an ambiguity in the application of this equations to the
case of one uniformly moving charge. Really, a simple charge translation
in space produces alterations of field components. Nevertheless, they can
not be treated in terms of Maxwell's displacement current. Strictly
speaking, in this case all Maxwell's displacement currents proportional to
$\partial{\bf E}/\partial t$ and $\partial{\bf B}/\partial t$ vanish from
eqs. (8)-(9). This statement can be reasoned by two different ways: ($i$)
$\partial{\bf E}/\partial t=0$ and $\partial{\bf B}/\partial t=0$, since
all field components of one uniformly moving charge are implicit
time-dependent functions (time enters as a unique parameter) so that from
the mathematical standpoint only total time derivative can be applied in
this case whereas partial time derivative turns out to be not adequate
(time and distance are not independent variables); ($ii$) a non-zero value
of $\partial{\bf E}/\partial t$ and $\partial{\bf B}/\partial t$  would
imply a local variation of fields in time independently of the charge
position and hence would imply the expansion of those local variations
through the propagation of electromagnetic waves. This would contradict
the fact that one uniformly moving charge does not radiate electromagnetic
field.

In this respect, it will be shown in this work that in a mathematically
consistent form of Maxwell-Lorentz set of equations all partial time
derivatives must be substituted by {\it total} ones. Only in this way all
ambiguities related to the application of Maxwell's displacement current
can be removed. On the other hand, it would imply a correct extension of
this concept to all quasistatic phenomena.

Thus, a mathematically rigorous interpretation of eqs. (8)-(9) in the case
of a charge moving with a constant velocity leads to the following
conclusion: in a free space out of a charge the value of $curl\,{\bf H}$
in eq. (8) is equal to zero:

\begin{equation}
curl\,{\bf H}=\frac{4\pi}{c}q{\bf V}\delta({\bf r}-{\bf
r}_q(t))
\end{equation}

On the other hand, the case of one uniformly moving charge has been
studied in detail and can be treated exactly in the framework of
Lorentz's transformations. Therefore, for any purpose one can apply exact
mathematical expressions for electric and magnetic fields and potentials
of a moving charge as follows [15]:
\begin{equation}
{\bf E}=q\frac{({\bf R}-R\mbox{\boldmath$\beta$})(1-\beta^2)}{(R-
\mbox{\boldmath$\beta$}{\bf R})^3},
\end{equation}
\begin{equation}
{\bf H}=\mbox{\boldmath$\beta$}\times{\bf E}
\end{equation}
where $\mbox{\boldmath$\beta$}={\bf V}/c$.

Thus, we arrive  at an important conclusion: generally speaking, the
value of $curl\,{\bf H}$ is {\it not} equal to zero in any point out of a
moving charge and takes a well-definite quantity:  \begin{equation}
curl\,{\bf H}=\frac{1}{c}({\bf V}\times{\bf E}).
\end{equation}

For instance, it gives immediately a non-zero value of $curl\,{\bf H}$
along the direction of motion ($X$-axis):
\begin{equation}
curl_x\,{\bf H}(x>x_q)=q\frac{2\beta(1-\beta^2)}{(1-\beta)^3(x-x_q)^3}.
\end{equation}

The conflict with the previous statement of equation (11) is inevitable. In
order to obtain symmetry between the set of field equations (6)-(9) and
their solutions one would expect an additional term like that considered
in (14). As it will be shown, this assumption resembles in
many respects the ingenious idea of Maxwell about the displacement current
that had revealed a profound symmetry and interrelation between the
electric and magnetic fields.

From the phenomenological point of view, the additional term could be
understood in the following manner. Really, every charge during its motion
changes the electric field flux through some fixed surface $S$ bounded by
the contour $C$. According to the Gauss' theorem, the same rate of
electric flux through $S$ might be produced from any other point of space.
It would correspond to some effective charge moving at the instant with a
certain velocity. In terms of Stokes' theorem this fact takes a clear
form. The circulation of the magnetic field is related with the amount of
the electric current passing through the surface of integration. Since all
surfaces enclosed by the contour $C$ are equivalent in accordance with the
conditions of Stokes' theorem then one could reasonably assume the
existence of {\it effective current} in the space out of a charge that
would make the same contribution into integral:
\begin{equation}
\oint\limits_{C}{\bf H}\,d{\bf l}=\int\limits_{S'}curl\,{\bf
H}\,d{\bf S}=\frac{4\pi}{c}\int\limits_{S}{\bf j}d{\bf S}
\end{equation}
where $S'$ and $S$ are equivalent surfaces bounded by $C$.

In the case of uniformly moving charge, this {\it effective current} play
the same role as the displacement current introduced by Maxwell for
non-steady processes. As it is well-known, the introduction of Maxwell's
displacement current was made on the base of following formal reasoning. In
order to make the equation (3) consistent with the electric charge
conservation law in form of continuity equation (5), Maxwell supplemented
(3) with an additional term. However, for stationary processes this term
disappears and the equation (3) becomes consistent only with closed (or
going off to infinity) currents:
\begin{equation}
div\,{\bf j}=0.
\end{equation}

On the other hand, this is a direct consequence of the continuity equation
(5) in any stationary state when all magnitudes must be considered as
implicit time-dependent functions. Thereby, we meet here another
difficulty of Lorentz's equations: uniform movement of a single charged
particle (as an example of {\it open steady current}), generally speaking,
does not satisfy the limitation imposed by equation (17). It implies some
additional term to be taken into consideration in (17) to fulfill the
Maxwell's condition on the circuital character of the total current. A
detailed analysis of the matter will be given in the next section.

Let us make a final remark about the following denomination that will be
adopted throughout this work. Further, we shall distinguish so-called
quasistatic fields among a more general class of common time-varying
fields. The study of steady processes in the original Maxwell's theory is
reduced to static fields when there is no change with time. It is possible
to overcome that limitation if one contrasts field alterations with time
in the case of uniformly moving charge from usual time-dependent fields.
As a matter of fact, it occupies an intermediate place between purely
static and common non-steady processes since all field values keep an
implicit time-dependence if the motion is uniform. This is reason to treat
such a case as quasistatic (or quasistationary) by definition.

\section{Balance equation and alternative form of Maxwell-Lorentz
equations for a single charged particle}

As it was pointed out in the {\it Introduction}, one of the principle
problems in the original Maxwell's theory was the relation between the
charge and matter. In fact, for many years Maxwell was adhered to the
Faraday's idea that the charge is a sort of field manifestation and at
first time  directed all his efforts in creating an unified theory of
matter as a field theory. However, further Maxwell adopted another
position in this respect by clear distinguishing between fields and matter
as two supplementary conceptions. Since then a charge had been entirely
attributed to the field and therefore opposed to matter. Finally, this
ambiguous approach constituted the foundation of so-called ``{\it
operative interpretation}" of Maxwell's field theory without any profound
assumptions on the charge nature and its relation with matter. In any case
it may be explained by the questionable status of many experimental data
about the charge at the middle of 19'th century. Fascinated by Faraday's
idea of tubes of field, Maxwell adopted this analogy also for electric and
magnetic fields as well as for electric charge flow (conduction currents).
As a consequence, in accordance with hydrodynamics language, the
continuity equation was accepted as valid to express the hypothesis that a
net sum of electric charges could not be annihilated. In this case, the
continuity equation reproduces the charge conservation law in the given
fixed volume ${\cal V}$:
\begin{equation}
\frac{d{\cal Q}}{dt}=\int\limits_{\cal
V}\Biggl(\frac{\partial\varrho}{\partial t}+div\,{\bf j}\Biggr)d{\cal V}=0
\end{equation}
or in form of differential equation:
\begin{equation}
\frac{\partial\varrho}{\partial t}+div\,{\bf j}=0\qquad\qquad
\Biggl(\frac{d{\cal Q}}{dt}=0\Biggr).
\end{equation}

It should be remarked that eq. (19) describes exclusively the conservation
but not the change of the amount of charge (or matter) in the given volume
${\cal V}$ as one can meet in some  or monographs that sometimes
make no distinction between this two aspects. If one wants to describe the
change of something in given volume ${\cal V}$, must replace the equation
(18) by a balance equation (see, e.g. [16]):
\begin{equation}
\frac{d{\cal Q}}{dt}=\frac{d}{dt}\int\limits_{\cal V}\varrho d{\cal V}=-
\oint\limits_S{\bf j}d{\bf S}=-\int\limits_{\cal V}div\,{\bf j}\,d{\cal V}.
\end{equation}
Here {\bf j} is a total current of electric charges through a surface $S$
that bounds the given volume ${\cal V}$. In mathematical language common to
all physical theories it means that the rate of increase in the total
quantity of electrostatic charge within any fixed volume is equal to the
excess of the influx over the efflux of current through a closed surface
$S$. On contracting the surface to an infinitesimal sphere around a point
one can arrive at the differential equation [16]:
\begin{equation}
\frac{d\varrho}{dt}+div\,{\bf j}=0\qquad\qquad \Biggl(\frac{d{\cal Q}}{
dt}\neq 0\Biggr).
\end{equation}

The balance equation (21) includes the continuity equation (19) as a
particular case when amount of something (charge or matter) maintains
constant in ${\cal V}$ with the course of time. In the previous section we
remarked that a single charge in motion, generally speaking, could not be
treated in terms of continuity equation (19) unless one is using so-called
{\it  slow motion approximation} [17]. By that condition is understood to
be a motion of charges in a certain limited region which charges do not
leave during the time of observation. Hence the equation (17) for one
uniformly moving charge is valid only under this condition. In more
general case, when a particle may leave the volume and thereby, violate
locally the charge conservation, one should be addressed to the balance
equation (21). One simple method is to prove it on an example of
point-charge moving with a constant velocity. In particular, the charge
density is assumed to have implicit time dependence as follows:
\begin{equation}
\varrho({\bf r},{\bf r}_q(t))=q\delta({\bf r}-{\bf r}_q(t))
\end{equation}
where ${\bf r}_q(t)$ is the coordinate of the charge at an instant $t$.

It is easy to show that the total density derivative in respect to time
will contain only the convection term, since the time enters in eq. (22)
as parameter $(\partial\varrho/\partial t=0)$:
\begin{equation}
\frac{d\varrho}{dt}=\frac{\partial\varrho}{\partial t}+
\Biggl(\left\{\frac{d}{dt}({\bf r}-{\bf
r}_q(t))\right\}\cdot\,grad\,\varrho\Biggr)=-({\bf V}\cdot\,grad\,\varrho)
\end{equation}
where ${\bf V}=d{\bf r}_q/dt$ is the velocity of the charge at the instant
$t$.

Thus, the balance equation for a single particle is fulfilled directly:
\begin{equation}
-({\bf V}\cdot\,grad\,\varrho)+div(\varrho{\bf V})=-({\bf
V}\cdot\,grad\,\varrho)+({\bf V}\cdot\,grad\,\varrho)=0.
\end{equation}

In spite of disappointing ambiguity of the original charge conception, all
direct followers of Maxwell (Thompson, Heaviside, FitzGerald and Pointing)
did not tried to make any additional assumptions on the relationship
between matter and charge. At first place, they were interested to
establish a true relation between a charge in uniform motion and a field
that surrounds it. The essence of their method was the Thompson's theory of
moving tubes of electric and magnetic forces. Without any explicit use of
continuity equation Thompson deduced that there must be displacement
currents in the space outside the sphere of electrostatically charged body
[7]. The further advance was effected by FitzGerald who had associated that
conclusion with the Maxwell's position in respect to circuital character
of the total electric current. In other words, the total current of
uniformly moving electrostatically charged body was to be composed of the
displacement current in outer space and the motion of body
itself so that the following condition might hold:
\begin{equation}
div({\bf j}_{\tt cond}+{\bf j}_{\tt disp})=0
\end{equation}
where ${\bf j}_{\tt cond}$ and ${\bf j}_{\tt disp}$ are conduction and
displacement currents respectively.

Here we shall not enter in details about the results obtained by Thompson
and et. al. but only mention that their approach for the first time
demonstrated the fundamental role of the displacement current (which was
entirely due to the translation of a charge and not to the explicit
time-varying processes) in the interrelation between quasistatic electric
and magnetic fields. The situation was so peculiar that Heaviside himself
insisted on the necessity to supplement Maxwell's equations with an
additional term. He assured that this term was so in the spirit of the
Maxwell's theory that if somebody had proposed this modification to
Maxwell, he would accepted it without any hesitation. From this position,
it seems that only balance equation in the case of uniformly moving charge
may be in accordance with the general Maxwell's condition (25) whereas the
continuity equation leads to the limited condition imposed by (17).

Really, we can rewrite the eq. (21) in the form of eq. (25) taking into
account the following denomination:
\begin{equation}
div\,{\bf j}_{\tt
disp}=\frac{d\varrho}{dt}=\frac{d}{dt}\left(\frac{1}{4\pi}\,div\,{\bf
E}\right)=div\left(\frac{1}{4\pi}\frac{d{\bf E}}{dt}\right)
\end{equation}

It may be easily verified that two operations $div$ and $\frac{d}{dt}$ are
completely interchangeable (for $div\,{\bf V}=0$) in (26). Thus, in a
particular case of an arbitrary moving charge when one can disregard its
size, the Maxwell's condition on total displacement current takes the
following form (see for the sake of comparison the formula (23)):
\begin{equation}
div\,{\bf j}_{\tt disp}=\frac{1}{4\pi}div\left\{\frac{\partial{\bf
E}}{\partial t}- ({\bf V}\cdot\nabla){\bf E}\right\} \end{equation}

So far we make use of the formal mathematical approach without any
physical interpretation. More specific, in calculating the full derivative
of {\bf E} in respect to time, the convection term in (27) must be
considered as quasistationary (fixed time condition) in agreement with the
definition of partial derivatives. In correct mathematical language it
means as if all field alterations produced by a simple charge translation
occur instantaneously (at same instant of time) {\it in every space
point}. This interpretation has no precedents in the classical
electrodynamics for the case of arbitrary moving charge whereas for an
uniform movement this quasistationary description is the {\it only
possible} formalism.

In this connection, recently, a validity of so-called {\it quasistationary
approximation} for, at least, one part of field quantities every instant
of time has been demonstrated in [12]. It is understood to be a direct
consequence of the mathematical condition on the continuity of general
solutions of the set of Maxwell's equations. In particular, for uniformly
moving charge, this approximation can be extended to all field values.
Indeed, in this case, all field alterations due to charge translation in
space occur instantaneously in every point, in accordance with the
conventional point of view. Thus it is assumed that all field quantities can
be separated into {\it two independent classes} with explicit $\{\}^*$ and
implicit $\{\}_0$ time dependence, respectively. The component ${\bf E}_0$
of the total electric field ${\bf E}$ in every point is understood to
depend {\it only} on the position of source at a given instant. In other
words, ${\bf E}_0$ is rigidly linked with the location of the charge. From
this point of view, the partial time derivative in (27) must be related
{\it only} with the explicit time dependent component ${\bf E}^*$ whereas
the convection derivative {\it only} with ${\bf E}_0$.
\begin{equation}
\frac{d{\bf E}}{dt}=\frac{\partial{\bf E}^*}{\partial t}-({\bf
V}\cdot\nabla){\bf E}_0;\qquad\mbox{where}\qquad {\bf E}={\bf E}^*({\bf
r},t) + {\bf E}_0({\bf R}(t))
\end{equation}
({\bf r} is a fixed distance from the origin of the reference system at
rest to the point of observation; ${\bf R}(t)={\bf r}-{\bf r}_q(t)$; ${\bf
r}_q(t)$ and ${\bf V}=\frac{d{\bf r}_q}{dt}$ are the distance and the
velocity of an electric charge.

 In spite of this
apparent relationship with the results developed in [12], they have been
advocated here exclusively for justifying the formal mathematical use
of the total time derivative in form of (27). As a matter of fact, the
separation of total field quantities into ${\bf E}_0$ and ${\bf E}^*$ is
obtained here independently and turns out to be the exact mathematical
result related to the existence of two non-reducible (independent) parts
in the expression of total time derivative. However, later on, we shall
use same denomination of field values in this two parts of (28), always
taking into account respective additional conditions. Only in the last
section we shall apply explicit separation of fields in basic equations in
order to find respective gauge transformations.

Thus, a general expression of full displacement current is then given by
the formulae:  \begin{equation} {\bf j}_{\tt
displ}=\frac{1}{4\pi}\frac{\partial{\bf E}}{\partial t}-
\frac{1}{4\pi}({\bf V}\cdot\nabla){\bf E}.
\end{equation}

Our initial aim was to find a reasonable form for the Maxwell's circuital
condition (25) that would allowed to relate field alterations in free space
produced by a single-moving charge with the Maxwell's conception of
displacement current. From the standpoint of conventional classical
electrodynamics, the first term represents the famous Maxwell's
displacement current appearing only in not-steady processes whereas the
second one could be understood as quasistationary term due to a simple
charge translation in space. Further, it would be convenient to denominate
this term as ``{\it convection displacement current}"

The above results motivate an important extension of displacement current
concept in the entire spirit of the Maxwellian original electromagnetic
theory. First, it postulates the circuital character of the total electric
current. Second, it permits to fulfill the circuital condition for
not-steady as well as for steady processes (static and quasistatic
fields), contrary to the conventional approach. Thus, we arrive
independently to Heaviside's intention of supplementing Maxwell's equations
with additional terms. As the last step before doing it, let us give an
equivalent mathematical expression of the convection displacement current
(in the case of single charged particle):
\begin{equation}
\frac{1}{c}({\bf V}\cdot\nabla){\bf E}=\frac{1}{c}\,{\bf V}\,div\,{\bf E}-
\frac{1}{c}\,curl\,({\bf V}\times{\bf E}).
\end{equation}

Accordingly, for our purpose we need to remind that in the right part of
eq. (8) the total current (${\bf j}_{\tt (tot)}={\bf j}_{\tt (cond)}+{\bf
j}_{\tt (disp)}$ must be considered like this:
\begin{equation}
curl\,{\bf H}=\frac{4\pi}{c}q{\bf V}\delta({\bf R}(t))+
\frac{1}{c}\frac{\partial {\bf E}}{\partial t}-
\frac{1}{c}{\bf V}\,div\,{\bf E}+
\frac{1}{c}curl({\bf V}\times{\bf E}).
\end{equation}

For completeness it is now necessary to assume that a moving charge may
also possess a certain quantity of magnetic moment. Despite the
difference between electric and magnetic charge conception we can usefully
limit our consideration with the magnetic analogy of balance equation.
This fortunate circumstance allows the treatment of eq. (9) in the same
way as (8):
\begin{equation}
curl\,{\bf E}=-\frac{1}{c}\frac{\partial {\bf B}}{\partial t}
+\frac{1}{c}({\bf V}\cdot\nabla){\bf B}= -\frac{1}{c}\frac{\partial {\bf
B}}{\partial t}+\frac{1}{c} {\bf V}\,div\,{\bf B}-\frac{1}{c}curl({\bf
V}\times{\bf B}).  \end{equation}

Turning to the previous section where some inconsistencies of Lorentz's
theory has been exposed, we see now that $curl\,{\bf H}$ is defined by
(31) in every point out of a charge in the expected manner (see eq.(14)).
As a final remark, eqs. (31,32) can be regarded as an alternative form of
Maxwell-Lorentz system of field equations (6)-(9). In next section they
will be analyzed on self-consistency and compared with modified
Maxwell-Hertz equations extended on one charge system.

\section{Alternative Maxwell-Hertz equations for one charge system}

Independently from Heaviside, the problem of modification of Maxwell's
equations for bodies in motion was posed by Hertz in his attempts to build
up a comprehensive and consistent electrodynamics [18]-[19]. A starting
point of that approach was the fundamental character of Faraday's law
of induction represented at first time by Maxwell in the form of integral
equations:
\begin{eqnarray}
&& \oint\limits_{C}{\bf H}\,d{\bf l}=\frac{4\pi}{c}\int\limits_{S}{\bf
j}\,d{\bf S}+\frac{1}{c}\frac{d}{dt}\int\limits_{S}{\bf E}\,d{\bf S}\\
&& \oint\limits_{C}{\bf E}\,d{\bf
l}=-\frac{1}{c}\frac{d}{dt}\int\limits_{S}{\bf B}\, d{\bf S}
\end{eqnarray}
where $C$ is a counter and $S$ is a surface bounded by $C$.

In qualitative physical language Faraday's observations had been expressed
in form of the following statement: {\it the effect of magnetic induction
in the circuit $C$ takes place everything with the change of the magnetic
flux through the surface $S$ independently weather it relates to the change
of intensity of adjacent magnet or occurs due to the relative motion}.
More over, Faraday established that the same effect was detected in a
circuit at rest as well as in that at motion. The latter fact provided a
principal base of Hertz's relativity principle which formulation resembles
in some aspects that of Einstein. In order to avoid details of Hertz's
original investigations [18]-[19], let us only note its sameness with
traditional non-relativistic treatment of integral form of Faraday's law
[20]. Namely, if the circuit $C$ is moving with a velocity {\bf v} in some
direction, the total time derivative in (33)-(34) must take into account
this motion (convection derivative) as well as the flux changes with time
at a point (partial time derivative) [20]:
\begin{equation}
\oint\limits_{C}{\bf E}\,d{\bf l}=-\frac{1}{c}\frac{d}{dt}\int\limits_{S}
{\bf B}\,d{\bf S}=-\frac{1}{c}\left\{\frac{\partial}{\partial t}+
({\bf v}\cdot\nabla)\right\}\int\limits_{S}{\bf B}\,d{\bf S}
\end{equation}
where $C$ and $S$ are any circuit and surface bounded by $C$,
respectively, moving together with a medium.

This approach is valid only for non-relativistic consideration and leads
to Galilean field transformation [20]. It was just one of the problem
which forced the complete rejection of Hertz's theory. More over, there
was no alternative in the framework of the respective ether model used by
Hertz. Any motion of the ether relative to the material particles had not
been taken into account, so that moving bodies were regarded simply as
homogeneous portions of the medium distinguished only by special values of
electric and magnetic constants. Among the consequences of such
assumption, Hertz saw the necessity to move the surface of integration in
eqs. (33)-(34) at the same time with the moving medium. Thus, the
generation of magnetic (or electric) force within a moving dielectric was
calculated with implicit use of Galilean invariance in eq. (35) unless one
makes any additional assumption on the special character of
transformations in a moving frame of reference.

A persistent experimental search for direct proofs of the existence of the
hypothetical ether led to the conclusion that its motion was completely
undetectable. The discovery of the special relativity principle by
Einstein that abolished at all the mechanistic conception of ether, gave
rise to a serious objection to Hertz's approach. However, in the following
we shall see a way to avoid those difficulties in the framework of the
integral formulation of Maxwell's equation.

Let us now examine the case of a point source of electric and magnetic
fields.\footnote{It should be noted that from mathematical standpoint of
the developed approach there is no matter to have a detailed information
about the nature of field source. Indeed, it may be a single charge,
dipole, a constant magnetic moment etc that generates a quasistatic
component ${\bf E}_0$ (or ${\bf B}_0$)  of the total field {\bf E} (or
{\bf B}) in a given point and at an instant.} In order to abstain from the
use of moving counter $C$ and surface $S$ that implies a direct
application of the relativity principle, we limit our consideration to a
fixed region ($C$ and $S$ are at rest) whereas the source is moving
through a free space. Accordingly to Faraday's law, there must be an
electromotive force in counter $C$ due to the flux changes with time as
well as with the motion itself. Generally speaking, this formulation must
treat the time and convection derivatives simultaneously. In a strict
mathematical language the convection part of the total derivative is to be
considered at fixed time condition (or is to be calculated for every point
at the same time) and, therefore, depends only on the position and the
velocity of the point field source at an instant. We have previously
mentioned the validity of the so-called {\it quasistationary
approximation} for one part of field derivatives that in the case of
uniformly moving charge may be extended to all field quantities. In other
words, in the framework of the present integral approach a treatment of
the total time derivative can be justified as analogous to differential
form (28):
\begin{equation}
\frac{d\Phi}{dt}=\frac{\partial\Phi^*}{\partial t}-({\bf V}_s\cdot\nabla)
\Phi_0
\end{equation}
making use of the definitions:
\begin{equation}
\Phi_0^{(E)}=\int\limits_S {\bf E}_0({\bf r}-{\bf r}_s(t))\,d{\bf S}
\qquad \mbox{or}\qquad
\Phi_0^{(B)}=\int\limits_S {\bf B}_0({\bf r}-{\bf r}_s(t))\,d{\bf S}
\end{equation}
and
\begin{equation}
\Phi^{*(E)}=\int\limits_S {\bf E}^*({\bf r},t)\,d{\bf S}
\qquad \mbox{or}\qquad
\Phi^{*(B)}=\int\limits_S {\bf B}^*({\bf r},t)\,d{\bf S}
\end{equation}
where ${\bf r}$ is a fixed distance from the origin of the reference
system at rest to the point of observation; ${\bf r}_s(t)$ and ${\bf
V}_s=d{\bf r}_s/dt$ are distance and the velocity of the electric (or
magnetic) field source.

For the sake of simplicity we shall conserve in this section the same
denomination of field flux in two independent parts of total time
derivative (36) taking into account tacitly additional (fixed space and
fixed time) conditions, respectively, in the following expression:
\begin{equation}
\frac{d}{dt}\Phi=\left\{\frac{\partial}{\partial t}-({\bf V}_s\cdot\nabla)
\right\}\Phi.
\end{equation}

Using a well-known representation of the convection part in eq. (36) like
that:
\begin{equation}
({\bf V}\cdot\nabla)\int\limits_S {\bf E}\,d{\bf S}=\int
\limits_S {\bf V}\,div\,{\bf E}\,d{\bf S}-\int\limits_S curl({\bf V}
\times{\bf E})\,d{\bf S}
\end{equation}
we obtain an alternative form of Maxwell integral equations (33)-(34) for
a moving electric charge in the reference system at rest:
\begin{eqnarray}
&& \oint\limits_C {\bf H}\,d{\bf l}=\frac{4\pi}{c}\int\limits_S {\bf
j}\,d{\bf S}+\frac{1}{c}\int\limits_S
\left\{\frac{\partial{\bf E}}{\partial t}-{\bf V}\,div\,{\bf E}+
curl({\bf V}\times{\bf E})\right\}d{\bf S}\\
&& \oint\limits_C {\bf E}\,d{\bf l}=-\frac{1}{c}\int\limits_S
\left\{\frac{\partial{\bf B}}{\partial t}+curl({\bf V}\times{\bf B})
\right\}d{\bf S}.
\end{eqnarray}
Likewise (31)-(32), it is also assumed a single charge to have additionally
a certain amount of a constant magnetic moment, i.e. to be a source of
electric and magnetic fields at the same time.

Before passing over to the more general consideration of a large number of
sources, we want to put attention to the most compact alternative
differential form of Maxwell-Hertz equations in the reference system at
rest:
\begin{eqnarray}
&& div\,{\bf E}=4\pi\varrho\\
&& div\,{\bf B}=0\\
&& curl\,{\bf H}=\frac{4\pi}{c}\varrho{\bf V}+\frac{1}{c}
\frac{d{\bf E}}{dt}\\
&& curl\,{\bf E}=-\frac{1}{c}\frac{d{\bf B}}{dt}
\end{eqnarray}
where the total time derivative of any vector field value {\bf E} (or {\bf
B}) can be calculated by the following rule:
\begin{equation}
\frac{d{\bf E}}{dt}=\frac{\partial{\bf E}}{\partial t}-({\bf
V}\cdot\nabla){\bf E}.
\end{equation}

For the first time the above-mentioned form (43)-(46) was adopted by Hertz
for electrodynamics of bodies in motion [18]-[19]. The only difference
consisted in the definition of the total time derivative that
corresponded to the moving medium (see (35)). A substitution of partial
time derivative in (1)-(4) by total derivatives in respect to time in the
alternative field equations (43)-(46) expresses the idea developed in [12]
(and supported independently in the present approach) that a  full
solution of complete and consistent set of Maxwell's equations must be
formed at the same time  by {\it implicit and explicit time-dependent
terms}. As it was yet mentioned, this fact is taken into account
automatically in (43)-(46) by the simultaneous coexistence of two
independent and mutually supplementary parts of total time derivative.

On the other hand, this alternative form of field equations removes the
ambiguity related with the application of the Maxwell's displacement
current concept in the case of a charge moving with a constant velocity.
Any alterations of field components in space must be treated in this case
exclusively in terms of the convection part of the total time derivative
(47) whereas partial derivative in respect to time vanishes and hence is
not adequate mathematically that confirms the analysis of some
inconsistencies in the form of Maxwell-Lorentz equations given in the
previous sections.

There is no difficulty to extend this approach to many particle system,
assuming the validity of electrodynamics {\it superposition principle}.
This extension is important in order to find out weather the alternative
form of microscopic field equations contains the original (macroscopic)
Maxwell's theory as limit case. To do that one ought to have into account
all principal restrictions of Maxwell's equations (1)-(5) which have only
dealing with continuous and closed (or going off to infinity) conduction
currents. They also must be motionless as a whole (static tubes of charge
flow), admitting only the variation of current intensity.

Under these conditions, it is quite easy to see that total (macroscopic)
convection displacement current is canceled all by itself by summing up
all microscopic contributions:
\begin{equation}
\sum_{a}({\bf V}_a\cdot\nabla){\bf E}_a=0;\qquad\qquad
\sum_{a}({\bf V}_a\cdot\nabla){\bf B}_a=0.
\end{equation}

In other words, every additional term in (31)-(32) (as well as in
(41)-(42)) disappears and we obtain the original set of Maxwell's
macroscopic equations (1)-(4) for continuous and closed (or going off to
infinity) conduction currents as valid approximation.

To conclude this section we want to note that the set of eqs. (41)-(42)
can be called as modified Maxwell-Hertz equations extended to one charge
system. It is easy to see that in this form they are completely equivalent
to modified Maxwell-Lorentz equations (31)-(32) obtained with the use of
the balance equation. Thus, differential and integral approaches to extend
the original Maxwell's equations seems to give the same result.

\section{Relativistically invariant formulation of alternative field
equations}

Let us write once again the alternative form of Maxwell-Lorentz equations
explicitly for a single moving particle that is a source of electric and
magnetic fields simultaneously:
\begin{eqnarray}
&& div\,{\bf E}=4\pi\varrho\\
&& div\,{\bf B}=0\\
&& curl\,{\bf B}=\frac{4\pi}{c}\varrho{\bf V}+\frac{1}{c}\left\{
\frac{\partial{\bf E}}{\partial t}-({\bf V}\cdot\nabla){\bf E}
\right\}\\
&& curl\,{\bf E}=-\frac{1}{c}\frac{\partial{\bf B}}{\partial t}+
\frac{1}{c}({\bf V}\cdot\nabla){\bf B}
\end{eqnarray}
at the same time with the balance equation:
\begin{equation}
\frac{d\varrho}{dt}+div\,\varrho{\bf V}=0.
\end{equation}

In the second section we already mentioned the necessity of a constant
taking into account respective additional conditions for each part of
total time derivative in (45)-(46). In other words, it corresponds to the
separation within the total values into two independent components. An
explicit use of this conditions in the basic field equations (51)-(52) can
be represented as follows:  \begin{eqnarray} && curl\,{\bf
B}=\frac{4\pi}{c}\varrho{\bf V}+\frac{1}{c}\left\{ \frac{\partial{\bf
E}^*}{\partial t}-({\bf V}\cdot\nabla){\bf E}_0\right\}\\ && curl\,{\bf
E}=-\frac{1}{c}\frac{\partial{\bf B}^*}{\partial t}+ \frac{1}{c}({\bf
V}\cdot\nabla){\bf B}_0 \end{eqnarray} where the total field values are
compound by two independent parts:  \begin{eqnarray} && {\bf E}={\bf
E}_0+{\bf E}^*={\bf E}_0({\bf r}-{\bf r}_q(t))+ {\bf E}^*({\bf r},t)\\ &&
{\bf B}={\bf B}_0+{\bf B}^*={\bf B}_0({\bf r}-{\bf r}_q(t))+ {\bf
B}^*({\bf r},t) \end{eqnarray} Here we note that quasistatic field
components ${\bf E}_0$ and ${\bf B}_0$ (further, we shall distinguish them
by putting subindex ``0") depend only on the point observation and on the
source position at an instant whereas time varying-fields ${\bf E}^*$ and
${\bf B}^*$ depend explicitly on time in a fixed point. The separation
procedure may by similarly extended to the electric and magnetic
potentials introduced as:  \begin{equation} {\bf
E}=-grad\,\varphi;\qquad\qquad {\bf B}=curl\,{\bf A} \end{equation} where
\begin{equation}
\varphi=\varphi_0+\varphi^*\qquad\mbox{and}\qquad
{\bf A}={\bf A}_0+{\bf A}^*.
\end{equation}

The invariance of the basic equations of the classical electrodynamics
under Lorentz transformations demands the system to be covariant in form.
Although this requirement of covariance is satisfied by the conventional
representation of Maxwell-Lorentz equations (6)-(9), there are some
difficulties with application of Lorentz gauge condition to quasistatic
fields. Before discussing it, let us prove the invariance of
alternative Maxwell-Lorentz (or Maxwell-Hertz) equations in form of
(49)-(52). This system can be further simplified by introducing from (55)
a so-called electromotive force:
\begin{equation}
{\bf E}=-grad\,\varphi-\frac{1}{c}\frac{\partial{\bf A}^*}{\partial t}-
\frac{1}{c}({\bf V}\times{\bf B}_0).
\end{equation}
Substituting {\bf E} in (54) we must use separately quasistatic and
time-varying parts of total electromotive force (60) in the following way:
\begin{equation}
{\bf E}_0=-grad\,\varphi_0-\frac{1}{c}({\bf V}\times{\bf B}_0)\qquad
\mbox{and}\qquad
{\bf E}^*=-grad\,\varphi^*-\frac{1}{c}\frac{\partial{\bf A}^*}{\partial
t}.
\end{equation}
The eqs. (50) and (52) (or (55)) are satisfied automatically and we are
left with the two differential equations:
\begin{eqnarray}
&& \Delta{\bf A}=-\frac{4\pi}{c}\varrho{\bf V}+{\bf F}\\
&& \Delta\varphi=-4\pi\varrho+{\cal F}
\end{eqnarray}
where
\begin{eqnarray}
&& {\bf F}=grad\,div({\bf A}_0+{\bf A}^*)-\frac{1}{c}({\bf
V}\cdot\nabla) grad\,\varphi_0+\frac{1}{c}\frac{\partial}{\partial
t}(grad\,\varphi^*)- \frac{1}{c^2}\frac{\partial^2{\bf
A}^*}{\partial{t^2}}\\
&& {\cal F}=-\frac{1}{c}div\left(\frac{\partial{\bf
A}^*}{\partial t}\right).
\end{eqnarray}
The second term in (64) can be easily transformed using mathematical
operations of field theory:
\begin{equation}
({\bf V}\cdot\nabla)grad\,\varphi_0=grad({\bf V}\cdot grad\,\varphi_0)-
({\bf V}\times curl\,grad\,\varphi_0).
\end{equation}
Since $curl\,grad$ is always equal to zero, we can rewrite ${\bf F}$ in
new form:
\begin{equation}
{\bf F}=grad\left\{div\,{\bf A}_0-\frac{1}{c}({\bf V}\cdot
grad\,\varphi_0)+div\,{\bf A}^*+\frac{1}{c}
\frac{\partial\varphi^*}{\partial t}\right\}+
\frac{1}{c^2}\frac{\partial^2{\bf A}^*}{\partial t^2}.
\end{equation}

The principal feature of (67) consists in the fact that all quasistatic
and time-varying components of total electric and magnetic potentials
enter independently and, therefore can be characterized by respective
gauge conditions:
\begin{eqnarray}
&& div\,{\bf A}_0-\frac{1}{c}({\bf V}\cdot grad\,\varphi_0)=0\\
&& div\,{\bf A}^*+\frac{1}{c}\frac{\partial\varphi^*}{\partial t}=0.
\end{eqnarray}

As a result, quasistatic fields turn out to be related through the novel
gauge (68) which is relativistically invariant and contains a well-known
relationship between the components of electric and magnetic field
potentials of uniformly moving charge [15]:
\begin{equation}
{\bf A}_0=\frac{\bf V}{c}\varphi_0.
\end{equation}

Recall that in the common point of view, electric and magnetic
potentials of uniformly moving charge are implicit time-dependent
functions. In this respect, (68) can be regarded as an extension of (70)
to all quasistatic quantities of electromagnetic field. On the other hand,
a unique reliable way to obtain (70) was based on the use of Lorentz
transformation. Here, it should be specially stressed the essential role
of the so-called {\it convection displacement current} conception in
deducing (70). So far, only time-varying fields were interrelated
explicitly due to Maxwell's displacement current. If one considers any
stationary process, such connection between quasistatic fields was yet
impossible since all displacement currents vanish from the Maxwell-Lorentz
equations. Contrary to this, the fundamental symmetry between quasistatic
electric and magnetic fields is now based on the equivalent conception of
the {\it convection displacement current}.

Another important aspect of the present approach can be attributed to the
verification of the limited character of the Lorentz gauge that now is
applicable only to the time-varying field components. In fact, there are
some difficulties in the conventional electrodynamics concerning the
inconsistency of this gauge with quasistatic potentials. Actually, in the
framework of the traditional approach, the Lorentz gauge condition
\begin{equation}
div\,{\bf A}+\frac{1}{c}\frac{\partial\varphi}{\partial t}=0
\end{equation}
is assumed to be valid for total electric and magnetic potentials and is
considered sufficient to hold the Maxwell's equations invariant under
Lorentz transformation. However, in the quasistationary approximation, the
Lorentz condition in every frame of reference takes the form of so-called
radiation gauge [21]:
\begin{equation}
div\,{\bf A}=0.
\end{equation}

On the other hand, due to the relation (70) between electric and magnetic
quasistatic potentials (72) is not satisfied directly. To make (72)
consistent with (70) in the given frame, it is suitable to put an
additional condition on the electric potential so that we arrive to the
so-called Coulomb gauge:
\begin{equation}
div\,{\bf A}=0\qquad\mbox{and}\qquad \varphi=0.
\end{equation}
In mathematical language the invariance of quasistatic fields involves
stronger limitations than those imposed previously by Lorentz gauge.
Generally speaking, the conventional classical electrodynamics must admit
more than one invariance principle since every time we make a Lorentz
transformation, we need also simultaneously transform all physical
quantities in accordance with the Coulomb gauge (73). This problem was
widely discussed and in the language adopted in the general Lorentz group
theory, is known as {\it gauge dependent representation (or joint
representation)} of the Lorentz group [21]. In fact, it means an
additional non-relativistic adjustment of electric potential, every time
we change the frame of reference. This difficulty takes no place if we
introduce the entirely relativistic gauge (68) for quasistationary
potentials.

A rigorous consideration of (62)-(65) gives another important conclusion:
simultaneous application of two independent gauge transformations
(68)-(69) decomposes the initial set (49)-(52) into two uncoupled pairs of
differential equations, namely:
\begin{eqnarray}
&& \Delta{\bf A}_0=-\frac{4\pi}{c}\varrho{\bf V}\\
&& \Delta\varphi_0=-4\pi\varrho
\end{eqnarray}
at the same time with the homogeneous wave equations:
\begin{eqnarray}
&& \Delta{\bf A}^*-\frac{1}{c^2}\frac{\partial^2{\bf A}^*}{\partial t^2}
=0\\
&& \Delta\varphi^*-\frac{1}{c^2}\frac{\partial^2\varphi^*}{\partial t^2}
=0.
\end{eqnarray}
Likewise (70), Poisson's second order differential equations (74)-(75) for
electric and magnetic potentials are in agreement with the conventional
approach in the steady state and can be
considered as valid extension to all quasistatic potentials.

The general solution, as one would expect, satisfies a pair of uncoupled
inhomogeneous D'Alembert's equations that can be verified by summing up
(74)-(75) and (76)-(77) (here we omit premeditatedly all boundary
conditions for the sake of simplicity):
\begin{eqnarray}
&& \Delta{\bf A}-\frac{1}{c^2}\frac{\partial^2{\bf A}}{\partial t^2}
=-\frac{4\pi}{c}\varrho{\bf V}\\
&& \Delta\varphi-\frac{1}{c^2}\frac{\partial^2\varphi}{\partial t^2}
=-4\pi\varrho
\end{eqnarray}
where the total values ${\bf A}$ and $\varphi$ are defined by (59).

The same result has been obtained in [12] independently from the analysis
of value boundary conditions for inhomogeneous D'Alembert's equations. As
a matter of fact, there has been shown that mathematically complete
general solution of Maxwell's equations must be written as a linear
combination of two non-reducible functions with implicit and explicit
time-dependence. Additionally, the present approach demonstrates the
invariance of (78)-(79) and therefore (49)-(52), if and only if two gauge
conditions (68)-(69) are satisfied by respective components of the total
field values.

To conclude this section, some remarks must be made concerning the
empirical and axiomatic status of Lorentz force conception in the electron
theory formulated by Lorentz. As we mentioned in {\it Introduction}, for
the first time an explicit formula for mechanical force acting on a
moving charge had been independently obtained in theoretical investigation
by Thompson and Heaviside. More over, in the first version of Maxwell's
theory published by the name ``{\it On Physical Lines of Force}"
(1861-1862) there was already admitted an unified character of a full
electromotive force in the conductor at motion by describing it as [22,4]:
\begin{equation}
{\bf E}=-\nabla\varphi-\frac{1}{c}\frac{\partial{\bf A}}{\partial t}+
\frac{1}{c}({\bf V}\times{\bf B})
\end{equation}
where the first term is the electrostatic force, the second one is the
force of magnetic induction and the third one is the force of
electromagnetic induction due to the conductor motion. It may be remarked
here that later investigators began to distinguish between the electric
force in a moving body and the electric force in the ether through which
the body is moving and as a result, did not consider $\frac{1}{c}({\bf
V}\times{\bf B})$ as a full-value part of electric field that afterwards
was argued by Hertz. This distinction had been profound by Lorentz in his
electron theory and was tightly related to the special status of the
Lorentz force conception. It also can be noted in the way how it forms
part the formalism of the conventional field theory. Really, the
mathematical form of (10) should be contrasted from the form of partial
differential equations (6)-(9) that make them weakly compatibles with (10)
contrary to that one would expect for the mathematical structure of a
complete and consistent system.

There are another formal arguments against the axiomatic and empirical
status of the Lorentz force. In classical mechanics one can find the very
similar notion of Coriolis force. It is also depends on velocity, not
position, and is not derivable from a potential. At the same time, being
always normal to {\bf V} it does not work and does not change the kinetic
energy of a particle. In spite of this sameness, an explicit expression
for the Coriolis force is deduced mathematically using the formalism of
classical mechanics. On the other hand, there is a simple way of
deriving of the Lorentz force, based on the transformation
of the electromagnetic field vectors and the components of the Minkovski
force.  Thus, the expression for the Lorentz force can be obtained in a
purely mathematical way from the general relations of the relativity
theory (see, for instance, [17]). All this remarks make the status of the
Lorentz force quite uncertain and pose the possibility of its
reconsideration. In this respect, the present approach allows to treat a
full electromotive force as completely unified conception that resembles
the formalism adopted in early stages of electromagnetic field theory. In
fact, the formula (60) used for the definition of full electromotive force
includes automatically a term responsible for the motion. A negative sign
of\quad $-\frac{1}{c}({\bf V}\times{\bf B}_0)$\quad can be easily
understood.  Contrary to (80) which corresponds to a moving medium and
fixed external magnetic field, in our particular case, a point of
observation (a site of a test charge) is fixed whereas a magnetic field is
``moving" with {\bf V}. In the reference system where magnetic field is at
``rest", the sign of the velocity is to be changed on the opposite and we
come in agreement with the direction of the Lorentz force. Therefore, we
can assume that for a consistent form of field equations there is no
necessity to supplement them by the equation of motion.

\section{Conclusion}

The above remarks motivate an important extension of the Maxwell's
concept of displacement current to all stationary electromagnetic
phenomena so that the fundamental symmetry between electric and magnetic
fields (including all quasistatic fields) can be understood now as a
consequence of the more general notion of displacement current. There are
other compelling reasons for seeking an alternative form of
Maxwell-Lorentz (as well as Maxwell-Hertz) equations that would contain
such an extension. This approach has demonstrated some advantages over the
conventional field description in eliminating a number of internal
inconsistencies from Lorentz's electrodynamics. One of them is concerning
the empirical and axiomatic status widely adopted in respect to the
Lorentz force conception that can be modified by assuming an unified
character of a total electromotive force. Thus, there is no more necessity
to supplement the set of Maxwell's equations by the equation of motion
since it  is taken into account automatically.

The rigorous solution of fields equations shows the existence of two
independent parts of field components correspond to longitudinal and
transverse modes. More fundamentally, the independence of two parts of
general solution must be attached on one hand to the implicit time
dependence of longitudinal solutions and, on the other, to the transverse
nature of explicitly time-varying fields. It arguments and corrects
the conventional point of view about the transverse character of the total
electromagnetic field and leads to the reformulation of the traditional
field concept in terms of the so-called {\it electrodynamics dualism
 concept:  simultaneous coexistence of instantaneous long-range
 (longitudinal) and Faraday-Maxwell short-rang (transverse) interactions}.
 More coherent and comprehensive analysis of this formulation and the
 compatibility of instantaneous action at a distance concept with the
 framework of relativistic classical electrodynamics can be found in [12].
 In this work we prefer to confine our consideration by alternative form
 of Maxwell-Lorentz and Maxwell-Hertz equations.

 Another aspect of the approach developed in this work relates to the
 deeper question of field gauge transformations. The existence of two
 independent types of fields (longitudinal and transverse) arrows us to
 determine independent gauge conditions for each other. The best
 way to end this article following A.O. Barut: ``... {\it Electrodynamics
 and the classical theory of fields remain very much alive and continue to
 be the source of inspiration for much of the modern research work in new
 physical theories}" [21].

\bigskip
\medskip

{\large Acknowledgments}

\bigskip

We are grateful to Dr. V. Dvoeglazov and Professor M. W. Evans for many
stimulating discussions.  Authors are indebted for financial support, R. S.-R., to the
Comunidad de Madrid, Spain, for the award of a Postgraduate Grant, A. Ch.,
to the Zacatecas University, M\'exico, for a Full Professor position.

\end{document}